\documentclass[doublecol]{epl2}
\usepackage{amssymb}

\title{Molecular dynamics simulations of the dipolar-induced formation
  of magnetic nanochains and nanorings}
\shorttitle{Molecular dynamics simulations of magnetic nanochains and nanorings}
\author{A. Hucht \and S. Buschmann \and P. Entel}

\institute{Department of Physics, University of
 Duisburg--Essen, 47048 Duisburg, Germany}

\abstract{
Iron, cobalt and nickel nanoparticles, grown in the gas phase, are
known to arrange in chains and bracelet-like rings due to the
long-range dipolar interaction between the ferromagnetic (or
super-paramagnetic) particles. We investigate the dynamics and
thermodynamics of such magnetic dipolar nanoparticles for low
densities using molecular dynamics simulations and analyze the
influence of temperature and external magnetic fields on two and three
dimensional systems. The obtained phase diagrams can be understood
by using simple energetic arguments. 
}

\begin{document}

\maketitle

The formation, structure and properties of nanoparticles grown in the
gas or liquid phase belongs to an active field of basic research, which
is of interest to future applications in different areas, from computer
technology to catalytic and biomedical applications (e.g., see
\cite{bhushan06}).  Magnetic nanoparticles show a variety of unusual
properties when compared to non-magnetic particles (for an overview,
see \cite{guet01}), which can often be attributed to the anisotropic
and long-range dipolar interaction between the particles.
For example, ferromagnetic and super-paramagnetic particles with low
mobility can be influenced by an external magnetic field, leading to
various arrangements and lattice structures both in experiments
\cite{WenWang99,Terheiden06}
and in computer simulations
\cite{HynninenDijkstra05a,HynninenDijkstra05b}.
On the other hand, in recent experiments on magnetic nanoparticles
with high mobility the formation of chains, rings and network-like
structures has been observed 
\cite{ButterBomans03,BlairKudrolli03,Knipping04,SnezhkoAransonKwok05,Huelser06,Salgueirino06}.
However, the influence of particle size, magnetic field, and
temperature on the chain formation is not well understood and is the
topic of this work. 

We consider a magnetic gas of $N$ identical, homogeneously magnetized
spherical particles with diameter $\sigma$, mass $m$, and magnetic moment 
$\vec{\mu}_i = \mu\hat{\vec{\mu}}_i$, located at position
$\vec{r}_i$. The particles are either enclosed in a three dimensional
($3d$) cube or restricted to a two dimensional ($2d$) substrate, both
with linear size $\tilde{L} = L/\sigma$. 
We use fixed boundary conditions, i.e.\  the particles are reflected by
the boundaries upon contact. The interaction potential of two
particles $i$ and $j$ at distance 
$\vec{r}_{ij} = \vec{r}_j - \vec{r}_i$ is  
$\mathcal{U}_{ij} = \mathcal{U}_{ij}^{\mathrm{d}}+\mathcal{U}_{ij}^{\mathrm{h}}$,
where $\mathcal{U}_{ij}^{\mathrm{d}}$ is the long-range anisotropic
dipolar interaction,
%
\begin{equation}
\mathcal{U}_{ij}^{\mathrm{d}} 
   = \frac{\mu_{0}\mu^{2}}{4\pi r_{ij}^{3}}
     \Big [ \hat{\vec{\mu}}_i \cdot\hat{\vec{\mu}}_j
   - 3 \left( \hat{\vec{\mu}}_i \cdot \hat{\vec{r}}_{ij} \right)
       \left( \hat{\vec{r}}_{ij} \cdot \hat{\vec{\mu}}_j \right)
     \! \Big ]
\label{eq:Uijd}
%
\end{equation}
%
with $\hat{\vec{r}}_{ij} = \vec{r}_{ij}/r_{ij}$, and 
$\mathcal{U}_{ij}^{\mathrm{h}}$ 
is a sufficiently rigid isotropic hard sphere interaction. Within this
work we chose
%
\begin{equation}
%
\mathcal{U}_{ij}^{\mathrm{h}}
   = \frac{\mu_{0}\mu^{2}}{16\pi\sigma^{3}}
     \left( \frac{\sigma}{r_{ij}} \right)^{\! 24},
\label{eq:UijHS}
%
\end{equation}
%
which leads for $N = 2$ particles to a ``head to tail'' ground
state configuration with energy $-E_{0}$, where
%
\begin{equation}
%
E_{0} =\frac{\mu_{0}\mu^{2}}{2\pi\sigma^{3}}.
\label{eq:E0}
%
\end{equation}
%
The total potential energy of $N$ particles in an external magnetic
field $\vec{B}$ is thus given by
%
\begin{equation}
%
\mathcal{U} = \frac{1}{2}\sum_{i\neq j}^{N} \mathcal{U}_{ij}
            - \sum_{i=1}^{N} \vec{B} \cdot \vec{\mu}_{i}.
\label{eq:U}
%
\end{equation}
%

In the molecular dynamics simulation, we numerically solve the equations
of motion for the positions, $\vec{r}_i(t)$, and the orientations of the
magnetic moments, $\vec{\mu}_i(t)$, of the particles under the influence
of the force $\vec{F}_i = \vec{F}(\vec{r}_i,t)$ and the local magnetic
field $\vec{B}_i = \vec{B}(\vec{r}_i,t)$ given by
%
\begin{eqnarray}
%
\vec{F}_i = -\frac{\partial\mathcal{U}}{\partial\vec{r}_i},
   \qquad 
\vec{B}_i = -\frac{\partial\mathcal{U}}{\partial\vec{\mu}_i}.
%
\end{eqnarray}
%
As the magnetic moment $\vec{\mu}_{i}$ is assumed to be pinned within the
particle by intrinsic anisotropies, the local magnetic field
$\vec{B}_i$ will provide a torque 
$\vec{M}_i=\vec{\mu}_i \times \vec{B}_i$
to the particle. The resulting equations of motion have the 
form
%
\begin{eqnarray}
%
\dot{\vec{p}}_i & = & m\ddot{\vec{r}}_i = \vec{F}_i,\\
\dot{\vec{L}}_i & = & I\dot{\vec{\omega}}_i 
                 = \vec{M}_i = \vec{\mu}_i \times \vec{B}_i,\\
\dot{\vec{\mu}}_i & = & \vec{\omega}_i \times \vec{\mu}_i,
%
\end{eqnarray}
%
were $\vec{p}_i$ is the momentum, $I = m\sigma^{2}/10$ the moment of
inertia, and $\vec{\omega}_i$ the angular velocity  of the $i$th
particle.

The equations of motion are integrated using the velocity Verlet
algorithm \cite{Verlet67}.
The molecular dynamics simulations can be performed within the 
micro-canonical ensemble with constant total energy,
$E^{\mathrm{tot}} = E^{\mathrm{kin}} + \mathcal{U} = \mathrm{const.}$,
and also within the canonical ensemble with constant thermal energy,
\begin{equation} \label{eq:Etherm}
  E^{\mathrm{{therm}}} = \frac{f}{2} N k_{\mathrm{B}}T = \left \langle 
  E^{\mathrm{{kin}}} \right \rangle = \mathrm{const.},
\end{equation}
where each particle has $f = 6$ degrees of freedom. In this case the
kinetic energy $E^{\mathrm{{kin}}}$ is adjusted using the Andersen
thermostat \cite{Andersen79}. 

\begin{figure} 
  \centerline{
    \includegraphics[bb=13bp 16bp 390bp 390bp,clip,scale=0.50]{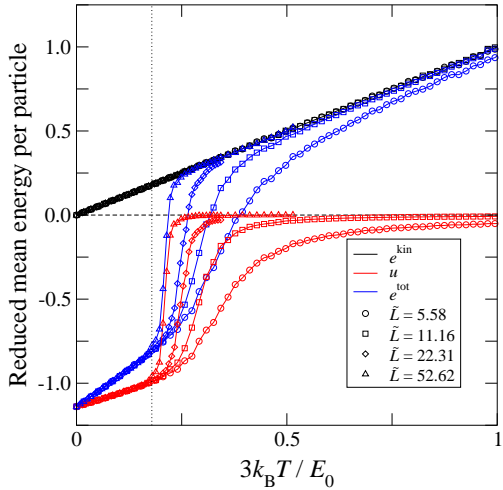}
  }
  \caption{\label{fig:1} (Color online): Kinetic energy
    $e^\mathrm{kin}$, potential energy $u$, and total energy
    $e^\mathrm{tot}$ of a $3d$ system with $N=9$ particles as function
    of temperature $T$ for different system sizes $\tilde{L}$, as
    obtained from the molecular dynamics simulation. The transition
    temperature $T_\mathrm{c}$ is shown as dotted line.  }
\end{figure}
%

\begin{figure} 
  \centerline{
    \includegraphics[bb=13bp 16bp 390bp 390bp,clip,scale=0.50]{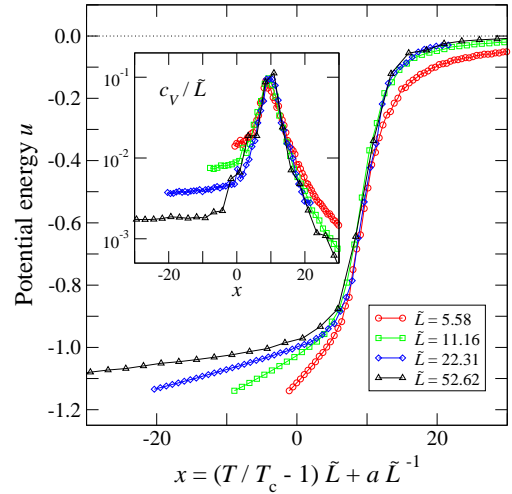}
  } 
  \caption{\label{fig:2} (Color online): Finite-size scaling plot of
    the potential energy $u$ and specific heat $c_{V}$ (inset) as
    function of the scaling variable $x$.
  }
\end{figure}
%
In the following, we present results of canonical simulations at fixed
volume $V$, particle number $N$, temperature $T$, and magnetic field
$\vec{B}$. We restrict $N$ to rather small values below 50 particles
in order to keep the statistical errors small. As we want to compare
our results to experiments done in the gas phase 
\cite{Knipping04,Huelser06},
we consider systems with very low particle densities and focus on the
case $L \rightarrow \infty$ at fixed $N$.
Figure~\ref{fig:1} shows the reduced kinetic energy 
$e^\mathrm{kin} = \langle E^{\mathrm{kin}} \rangle / E_0 N$, 
the reduced potential energy
$u              = \langle \mathcal{U}      \rangle / E_0 N$
and the reduced total energy
$e^\mathrm{tot} = \langle E^{\mathrm{tot}} \rangle / E_0 N$
versus reduced temperature for a system consisting of $N=9$ particles,
with different linear size $\tilde{L} = L/\sigma$ in three
dimensions. $N=9$ is just one particular choice, $N>9$ yields
similar results but larger statistical errors. It can be seen that the
potential energy $u$ develops a jump with growing system size $L$,
which results in a jump in the total energy $e^\mathrm{tot}$, as the
kinetic part is simply proportional to temperature in the canonical
ensemble (see eq. (\ref{eq:Etherm})). 
The $L$-dependence of $u$ stems from the
fact that in a finite simulation volume particles which leave the
bound state, chain or ring, are reflected back towards the bound
structure by the boundaries. Thus the bound state is more stable in
small volumes than in large or even infinite volume. This effect
broadens the width of the transition and shifts the effective
transition temperature $T_\mathrm{c}(L)$ to higher temperatures for
smaller $L$. 

In order to determine the nature of the transition as well as the
critical temperature $T_{\mathrm{c}}$, a finite-size scaling plot of
the potential energy $u(T)$ is shown in fig.~\ref{fig:2}, together
with the specific heat at constant volume and constant number of
particles, 
$c_{V} = \partial u / \partial T |_{V,N}$.
Using the scaling variable 
\begin{equation}
 x = \left( \frac{T}{T_{\mathrm{c}}} - 1 \right) \tilde{L} + \frac{a}{\tilde{L}},
\end{equation}
a data collapse both in $u(T)$ and in 
$c_{V}(T)$ can be achieved, leading to an estimation for the critical 
 temperature $T_{\mathrm{c}}$,
%
\begin{equation}
%
k_\mathrm{B} T_\mathrm{c} = 6.0(6) \times 10^{-2}  E_\mathrm{0}.
\label{eq:Tc}
%
\end{equation}
%
The constant $a$ describes corrections to scaling and has the
value $a = 25(3)$.  The fact that no rescaling is necessary in $u$
shows that the transition is of first order.

%
\begin{figure} 
   \centerline{ 
     \includegraphics[clip,scale=.9]{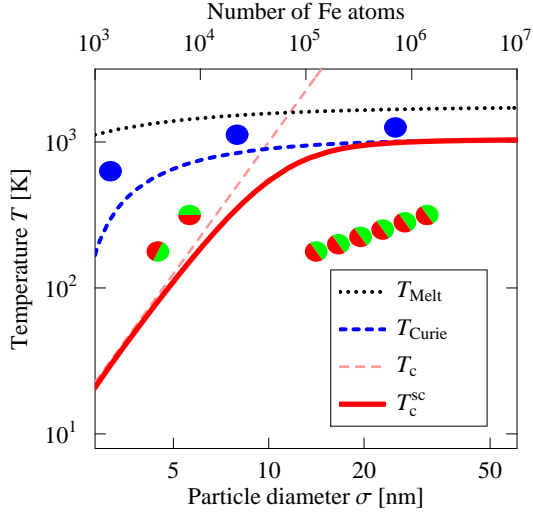}
                }
   \caption{\label{fig:3}
   (Color online): Phase diagram of magnetic nanoparticles
   as function of particle size $\sigma$ and temperature $T$. Shown are the
   melting temperature (dotted black), the Curie temperature (dashed blue), 
   and the chain building temperature, both with constant moment 
   $\mu$ (long dashed pink) as well as with temperature dependent 
   moment $\mu(T)$ (thick red), see text.
}
\end{figure}
%
In the following discussion we neglect quantum mechanical effects
which lead to, e.g., size-dependent magnetic moments in Fe particles
with less than 1000 atoms \cite{Billas93}.  In our model the ground
state energy $E_\mathrm{0}$ depends on the size $\sigma$ and magnetic
moment $\mu$ of the particle via eq.~(\ref{eq:E0}), leading to a
simple dependence of the critical temperature $T_\mathrm{c}$ on the
considered type of nanoparticles.  For instance, Fe nanoparticles with
a diameter of $\sigma \approx 6\,\textrm{nm}$ ($10^4$ atoms) have a
saturation magnetization of $\mu \approx 2.2 \times 10^4
\mu_\mathrm{B}$, where $\mu_\mathrm{B}$ is the Bohr magneton. The
critical temperature for the Fe nanoparticles follows from
eq.~(\ref{eq:Tc}) to $T_\mathrm{c} \approx 140\,\textrm{K}$.  A
resulting phase diagram for magnetic nanoparticles is depicted in
fig.~\ref{fig:3}, where $T_\mathrm{c}(\sigma)$ is shown as a dashed
line. As to lowest order, $T_\mathrm{c}(\sigma) \propto \sigma^3$, it
has to be compared to other characteristic temperatures like the Curie
temperature $T_\mathrm{Curie}$ and structural transition temperatures,
e.g., the melting temperature $T_\mathrm{Melt}$, in order to
determine the validity of the theory. Hence these two temperatures,
which are size dependent as well, are also depicted schematically for
typical magnetic systems. The finite-size form of $T_\mathrm{Melt}$
can be approximated to lowest order using energetic arguments to give
$T_\mathrm{Melt}(\sigma)/T_\mathrm{Melt}^\infty = 1 - \sigma_\mathrm{Melt} / \sigma$
with material specific constant $\sigma_\mathrm{Melt}$ \cite{Pawlow1909}.
In the case of the magnetic transition, standard finite-size scaling
theory gives, again to lowest order, a temperature shift of the form 
$T_\mathrm{Curie}(\sigma)/T_\mathrm{Curie}^\infty = 1 - 
(\sigma_\mathrm{Curie}/\sigma)^{1/\nu}$,
with the exponent of magnetic correlations $\nu$ and
constant $\sigma_\mathrm{Curie}$ \cite{Hendriksen93}. As the
magnetization $\mu$ vanishes at $T_\mathrm{Curie}$ as
$\mu(T)=\mu(0)(1-T/T_\mathrm{Curie})^\beta$ with critical exponent $\beta$,
and as $E_0$ is proportional to $\mu^2$ (eq.~(\ref{eq:E0})), we can
calculate a corrected $T_\mathrm{c}^\mathrm{sc}(\sigma)$ with
temperature dependent magnetic moments as solution of the
self-consistency equation 
\begin{equation} \label{eq:Tcsc}
\frac{T_\mathrm{c}^\mathrm{sc}(\sigma)}{T_\mathrm{c}(\sigma)} = 
\left( 1 - \frac{T_\mathrm{c}^\mathrm{sc}(\sigma)}{T_\mathrm{Curie}(\sigma)} \right)^{2\beta},
\end{equation}
shown in fig.~\ref{fig:3} as thick red line \cite{parameters}.
For small particles
$T_\mathrm{c}^\mathrm{sc}(\sigma) \approx T_\mathrm{c}(\sigma)$, as
the rhs.\ of eq.~(\ref{eq:Tcsc}) is approximately unity.
On the other hand, for big particles
$T_\mathrm{c}^\mathrm{sc}(\sigma) \approx T_\mathrm{Curie}(\sigma)$,
as now the lhs.\ of eq.~(\ref{eq:Tcsc}) is vanishing.
As a result, in the case of Fe nanoparticles, a phase transition from
ferromagnetic chains/rings to a ferromagnetic gas can occur for
particle sizes below approximately 20~nm, while for larger particles
the transition goes from ferromagnetic chains/rings to a paramagnetic gas. 

\begin{figure} 
   \centerline{ 
   \includegraphics[bb=13bp 16bp 390bp 390bp,clip,scale=0.50]{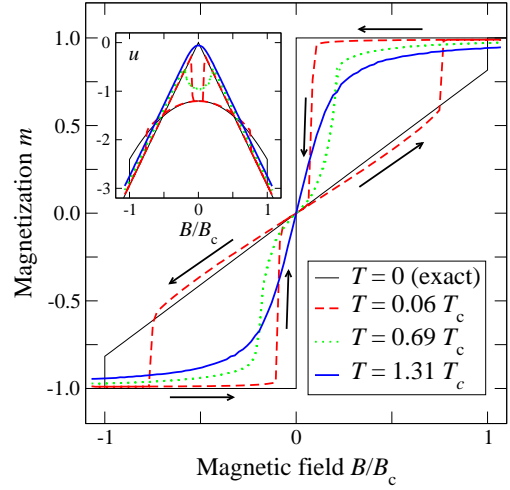}
   }
   \caption{\label{fig:4} (Color online): Magnetization $m(B)$ of
   $N=25$ particles confined to a $2d$ surface as a function of the
   magnetic field $B$ for different temperatures $T$. The arrows
   indicate the direction of the hysteresis loop. The corresponding
   mean reduced potential energy $u(B)$ is shown in the inset.  }
\end{figure}%
In the following we consider two and three dimensional systems in an
external magnetic field $\vec B$, which in the $2d$ case is oriented
perpendicular to the plane.
The field $B=|\vec{B}|$ allows us to analyze the stability of the
structures which are observed in the simulations -- namely a closed
ring, a chain and a lattice-like structure in the $2d$ case. Typical
hysteresis loops of a $2d$ system for different temperatures are shown
in fig.~\ref{fig:4}.
At low temperatures, $T=0.06~T_\mathrm{c}$, the system shows a broad
hysteresis loop.  Starting from the initial ring structure with the
zero field value, $m=0$, the magnetization grows linearly while the
moments of the particles start to turn into the direction of the
increasing field. The following jump to the saturation value is
related to the dissolving of the ring, because afterwards the
particles are unbonded, forming a lattice-like structure on the $2d$
surface. Then, the moments of the particles point into the direction
of the field, which gives rise to the saturation magnetization of the
system. With decreasing field, the magnetization remains saturated and
the particles stay in the lattice-like structure until this structure
becomes unstable due to thermal fluctuations. The jump down to the
linear part of the magnetization follows from the structural
rearrangement of the particles back to the ring structure.
With increasing temperature, the width of the hysteresis loop becomes
smaller and vanishes at $T \approx 0.69~T_\mathrm{c}$.
In the temperature range
$0.69~T_\mathrm{c} \lesssim T < T_\mathrm{c}$ the potential energy
$u(B)$ shows a cusp at $B_\mathrm{c}$ (see inset of fig.~\ref{fig:4}),
which vanishes at the critical temperature $T_\mathrm{c}$. Above
$T_\mathrm{c}$ the system is paramagnetic.


In the following, a simple model is presented, which, based on
analytical ground state arguments like the potential energy, is able
to explain qualitatively the numerical results in both $2d$ and $3d$
systems. For a large number of particles, $N \rightarrow \infty$, both
the ring and the chain structures can be treated equally regarding
their potential energy $u$. Due to the long-range nature of the
dipolar interaction, the corresponding result for the reduced
potential energy in the ground state is $u=-\zeta(3)$, where
$\zeta(s)$ denotes the Riemann Zeta function. 
In both $2d$ and $3d$ systems at $T < T_{\mathrm{c}}$ and $B=0$ the
particles form a ring, which is the most favorable state for $N \geq 4$
\cite{WenKun99}. In the presence of a magnetic field oriented
perpendicular to the system in $2d$, the equilibrium state is changed:
In the $3d$ case, the ring plane turns perpendicular to the field,
similar to the behavior of an antiferromagnetic spin systems in an
external magnetic field.
Furthermore, in both cases the magnetic moment $\vec{\mu}_{i}$ of each
particle turns into the field direction by an angle $\alpha(B)$,
leading to a potential energy per particle,
%
\begin{equation}
%
u(\alpha) = -\zeta(3) \left[
             \frac{3}{2}\cos^{2}(\alpha) - \frac{1}{2}
	     +\sqrt{6} \frac{B}{B_\mathrm{c}} \sin(\alpha) \right]
\label{eq:u(alpha)}
%
\end{equation}
%
with $B_{\mathrm{c}} = \sqrt{6}\zeta(3)E_{0}/\mu$.
The equilibrium state is characterized by the angle 
%
\begin{equation}
\alpha(B) = \arcsin \left( \sqrt{
                    \frac{2}{3} } \frac{B}{B_{\mathrm{c}}} \right),
\label{eq:alphanull}
\end{equation}
%
which follows from eq.~(\ref{eq:u(alpha)}) by requiring $\partial
u/\partial \alpha = 0$.  Using this dependency, we can express the
potential energy $u$ as a function of the magnetic field to get
%
\begin{equation}
%
u(B) = -\zeta(3) \left[ 1 + \left( \frac{B}{B_{\mathrm{c}}} \right) ^2 \right].
%
\end{equation}
%
This curve is shown as thin black line in the inset of fig.~\ref{fig:4}.

The ring structure remains stable as long as $B < B_{\mathrm{c}}$ and
the dipolar interaction is attractive. This changes at $B =
B_{\mathrm{c}}$, where the angle $\alpha$ approaches the critical
value
$\alpha(B_{\mathrm{c}}) = \arcsin(\sqrt{2/3}) \approx
54.74^{\circ}$. Here, the dipolar interaction becomes repulsive,
leading to a break up of the ring due to the mobility of the
particles
(note that $B_{\mathrm{c}}$ is independent of the size of the
particles; for saturated Fe spheres we find
$B_\mathrm{c}\approx 0.47~\mathrm{T}$).
In the $3d$ case, the particles rearrange in chains
aligned parallel to the field, which is not possible in the $2d$
case, as the particles are confined to the surface. Instead, they
rearrange in a hexagonal lattice-like structure on the surface, which is
stabilized due to the repulsive character of the dipolar interaction.
Now, both in three and in two dimensions, $u$ is mainly dominated by
the field energy being proportional to $-B$
(see inset in fig.~\ref{fig:4}).

\begin{figure}[t] 
   \centerline{ \includegraphics[scale=0.9]{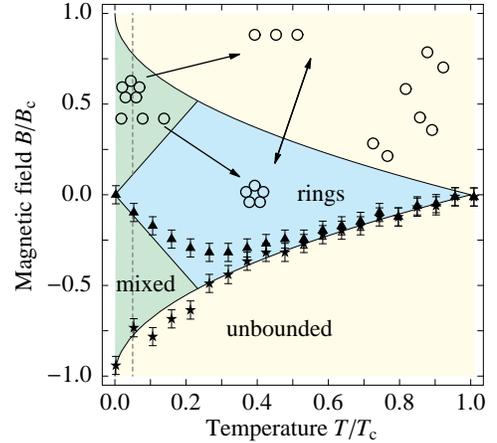} }
   \caption{\label{fig:5} (Color online) Phase diagram of the $2d$
   system. The arrows indicate the different structures of
   nanoparticles when crossing the phase boundaries. Stars mark the
   transition from the bonded to the unbonded phase, triangles the
   reverse transitions, as obtained from the molecular dynamics
   simulation. For clarity, the simulation data is only shown in the
   lower half plane. The dashed line marks the transition to the
   lattice-like structure with decreasing temperature.  }
\end{figure}
\begin{figure}[t] 
   \centerline{
      \includegraphics[scale=0.9]{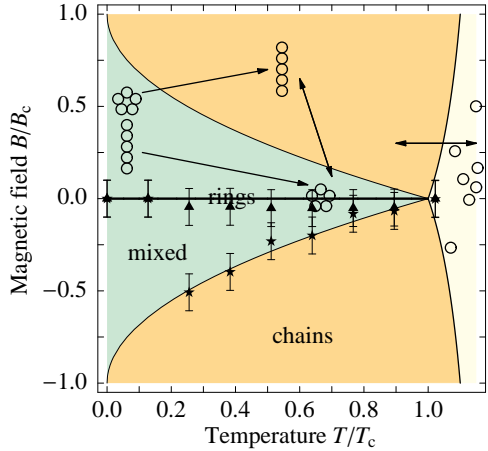}
   }
   \caption{\label{fig:6}
     (Color online) Phase diagram of the $3d$ system. The arrows indicate 
     the different structures of nanoparticles when crossing the phase 
     boundaries. Stars mark the transition from the ring to the chain 
     phase, triangles the reverse transitions, as obtained from the 
     molecular dynamics simulation. For clarity, the simulation data
     is only shown in the lower half plane.
   }
\end{figure}

The reduced magnetization $m$ parallel to the magnetic field can be
calculated for ring structures as a function of the field $B$,
which yields
%
\begin{eqnarray}
%
  m(\alpha)    & = & \frac{M(\alpha)}{M_{\mathrm{s}}}=\sin(\alpha),
  \label{eq:malpha}
  \\
  m(B) & = & \sqrt{\frac{2}{3}}\frac{B}{B_{c}},
  \label{eq:mb}
  %
\end{eqnarray}
%
where $M_{\mathrm{s}}$ is the saturation magnetization (thin black
line in fig.~\ref{fig:4}).

These ground state arguments can be expanded to describe the influence
of temperature on the system, by specifying the transitions between
the different states by purely energetic arguments. At finite
temperatures and non-vanishing magnetic fields, the ring can be broken
up by overcoming an energy barrier $\Delta u$ by thermal
activation. This energy barrier separates the ring and the
lattice-like structure in $2d$ and the ring and the chain structure in
$3d$ respectively. 
Basically the barrier height is given by
$\Delta u(B) = u(B_\mathrm{c}) - u(B)$,
which is the energy difference between the unstable ring and ring in
equilibrium in the presence of the field $B$,
(eq.~(\ref{eq:u(alpha)})). If the mean kinetic energy 
equals this barrier height, the ring becomes unstable and dissolves.
This and the equivalent consideration for the transition back to the
ring structure leads to
%
\begin{equation}
%
B^{\uparrow}_{\mathrm{c}}(T) = B_{\mathrm{c}}
   \left(1- \sqrt{\frac{T}{T_{\mathrm{c}}}} \right).
\label{eq:Bup}
%
\end{equation}
%
If the field exceeds the critical value
$B^{\uparrow}_{\mathrm{c}}(T)$, the ring dissolves into the
lattice-like structure due to the kinetic fluctuations and the
influence of the field. For fields lower than
%
\begin{equation}
%
B^{\downarrow}_{\mathrm{c}}(T)
   = B_{\mathrm{c}} \left(1+ \sqrt{\frac{3}{2}} \right)
     \frac{T}{T}_{\mathrm{c}},
\label{eq:Bdown}
%
\end{equation}
%
the dipolar particles flip back to the ring structure.
Figure~\ref{fig:5} shows the phase diagram of the $2d$ system using
the calculated phase boundaries, eq.~(\ref{eq:Bup}) and
eq.~(\ref{eq:Bdown}), together with the data obtained from the
molecular dynamics simulations. Both the lattice-like and ring
structure can coexist in the area which is limited by these phase
boundaries until the intersection is reached. If the temperature
exceeds this boundary, only one state can survive depending on the
field, leading to the loss of hysteresis.
The pure ring phase is limited by the phase boundaries until
$T_\mathrm{c}$ is reached.  If the critical temperature $T_\mathrm{c}$
is reached at zero field, the rings dissolve and the particles enter
the paramagnetic phase. For fields larger than
$B^{\uparrow}_{\mathrm{c}}(T)$, the particles are unbonded, whereas
the lattice-like structure is sustainable only at low temperatures,
$T \ll T_\mathrm{c}$. The lattice-like structure gets lost due to the
growing kinetic energy of the particles with increasing temperature.

The phase diagram of the $3d$ system is depicted in fig.~\ref{fig:6}.
We find the pure chain phase and a mixed phase where both rings and
chains can coexist. The pure ring phase is, in contrast to the $2d$
system, limited to $B=0$. Therefore, the area in which we can observe
hysteresis is larger than in $2d$ systems. At high temperatures,
$T>T_\mathrm{c}$, the system enters the paramagnetic phase, where the
particles are unbonded. However, if the field is switched on the chain
phase is again stabilized, because all particles are forced to point
into the same direction. This makes an arrangement parallel to the
field more favorable. At low temperatures, $T < T_{\mathrm{c}}$, and
high fields, $B > B^{\uparrow}_{\mathrm{c}}$, the particles can
arrange only in chains. If the field drops to zero, the chain closes
to rings provided that this bending is affordable. Otherwise, a short
chain will turn into the direction of the field without closing to a
ring. In this case, the system can go from the upper to the lower
mixed phase without entering the pure ring phase. This behavior
depends on the length of the chain because it is more favorable to
bend longer chains to closed rings than shorter ones. This shifts the
starting point of the pure ring phase from $T=0$ to higher
temperatures. 

In this paper we have derived a phase diagram of dipolar nanoparticles
as a function of particle size and temperature, explaining the
transition from magnetic chains and rings to a magnetic gas.
Furthermore, we have calculated the phase diagrams for two and three
dimensional systems as a function of temperature and external magnetic
fields by molecular dynamics simulations. The different regions of
stability explain qualitatively the collective behavior of magnetic
nanoparticles which has been observed in several experiments.

\acknowledgments
Financial support by the DFG (SFB 445) is acknowledged.


\begin{thebibliography}{15}
\bibitem{parameters}
In fig. \ref{fig:3} the following parameters are used:
$T_\mathrm{Melt}^\infty = 1743\,\mathrm{K}$,
$\sigma_\mathrm{Melt} = 1\,\mathrm{nm}$,
$T_\mathrm{Curie}^\infty = 1043\,\mathrm{K}$,
$a_\mathrm{Curie} = 2.5\,\mathrm{nm}$,
$\nu=0.7$, and $\beta=0.33$.
\end{thebibliography}
\end{document}